\documentclass[preprint,amssymb,nobibnotes,nofootinbib,aps,prc]{revtex4}
\usepackage{psfig}
\usepackage{graphicx}
\usepackage{bm}
\newcommand{\bea}{\begin{eqnarray}}
\newcommand{\eea}{\end{eqnarray}}
\newcommand{\be}{\begin{equation}}
\newcommand{\ee}{\end{equation}}
\newcommand{\bt}{\begin{tabular}}
\newcommand{\et}{\end{tabular}}
\newcommand{\Tr}{{\rm Tr}}
\newcommand{\no}{\nonumber}
\newcommand{\ovl}{\overline}

\newcommand{\Si}{ \mbox{\boldmath $\Sigma$}  }

\newcommand{\pa}{\partial}
\newcommand{\beas}{\begin{eqnarray*}}
\newcommand{\eeas}{\end{eqnarray*}}

\newcommand{\fr}{\frac}

\newcommand{\g}{\gamma}

\newcommand{\dg}{\dagger}
\newcommand{\La}{\Lambda}
\newcommand{\pam}{\partial_\mu}
\begin{document}

\title{Kaons and antikaons in asymmetric nuclear matter}

\author{Amruta Mishra}
\email{amruta@physics.iitd.ac.in}
\affiliation{Department of Physics,Indian Institute of Technology,Delhi,
New Delhi - 110 016, India}

\author{Stefan Schramm}
\email{schramm@th.physik.uni-frankfurt.de}
\affiliation{
Frankfurt Institute for Advanced Studies,
J.W. Goethe Universit\"at,
Ruth-Moufang-Str. 1, D-60438 Frankfurt am Main, Germany}

\author{W.~Greiner}
\affiliation{
Frankfurt Institute for Advanced Studies,
J.W. Goethe Universit\"at,
Ruth-Moufang-Str. 1, D-60438 Frankfurt am Main, Germany}

\begin{abstract}
The properties of kaons and antikaons and their modification in isospin
asymmetric nuclear matter are investigated using a chiral SU(3)
model. These isospin dependent medium effects are important for
asymmetric heavy ion collision experiments. In the present work,
the medium modifications of the energies of the kaons and
antikaons, within the asymmetric nuclear matter, arise due to the
interactions of kaons and antikaons with the nucleons and scalar
mesons. The values of the parameters in the model are obtained by
fitting the saturation properties of nuclear matter and
kaon-nucleon scattering lengths. The pion-nucleon scattering
lengths are also calculated within the chiral effective model and
compared with earlier results from the literature. The density
dependence of the isospin asymmetry is seen to be appreciable for
the kaon and antikaon optical potentials. This can be particularly
relevant for the future accelerator facility FAIR at GSI, where
experiments using neutron rich beams are planned to be used in the
study of compressed baryonic matter.

\end{abstract}


\maketitle

\section{Introduction}

The study of the properties of hadrons in hot and dense matter
is an important topic in strong interaction physics.
This subject has direct implications for heavy-ion collision
experiments, in the study of astrophysical compact objects (like
neutron stars) as well as in the early universe. The in-medium
properties of kaons have been investigated particularly because of
their relevance for neutron star phenomenology as well as
relativistic heavy-ion collisions. For example, in the interior of
a neutron star the attractive kaon-nucleon interaction might lead
to kaon condensation as originally suggested by Kaplan and Nelson
\cite{kaplan}.  The in-medium modification of kaon/antikaon
properties can be observed experimentally primarily in
relativistic nuclear collisions. Indeed, the experimental
\cite{FOPI,Laue99,kaosnew,Sturm01,Forster02} and theoretical
studies
\cite{lix,cmko,Li2001,Cass97,brat97,CB99,laura03,Effenber00,Aichelin,Fuchs}
of $K^\pm$ production in A+A collisions at SIS energies of 1-2
A$\cdot$GeV have shown that in-medium properties of kaons can be
connected to the collective flow pattern of $K^+$ mesons as well
as to the abundance and spectra of antikaons.

The theoretical research work on the topic of medium modification
of hadron properties was initiated by Brown and Rho \cite{brown}
who suggested that the modifications of hadron masses should scale
with the scalar quark condensate $\langle q\bar{q}\rangle$ at
finite baryon density.  The first attempts to extract the
antikaon-nucleus potential from the analysis of kaonic-atom data
were in favor of very strong attractive potentials of the order of
-150 to -200 MeV at normal nuclear matter density $\rho_0$
\cite{FGB94,Gal}.  However, more recent self-consistent
calculations based on a chiral Lagrangian
\cite{Lutz98,Lutz021,Lutz02,Oset00} or coupled-channel G-matrix
theory (within meson-exchange potentials) \cite{lauran} only
predicted moderate attraction with potential depths of -50 to -80
MeV at density $\rho_0$.

The problem with the antikaon potential at finite baryon density
is that the antikaon-nucleon amplitude in the isospin channel
$I=0$ is dominated by the $\Lambda(1405)$ resonant structure,
which in free space is only 27 MeV below the ${\bar K}N$
threshold. It is presently not clear if this physical resonance is
an excited state of a `strange' baryon or some short lived
molecular state that, for instance, can be modelled in
a coupled channel $T$-matrix scattering equation using a suitable
meson-baryon potential. Additionally, the coupling between the
${\bar K}N$ and $\pi Y$ ($Y=\Lambda,\Sigma$) channels is essential
to get the proper dynamical behavior in free space.
Correspondingly, the in-medium properties of the $\Lambda(1405)$,
such as its pole position and its width, which in turn strongly
influence the antikaon-nucleus optical potential, are very
sensitive to the many-body treatment of the medium effects.
Previous works have shown that a self-consistent treatment of the
$\bar{K}$ self energy has a strong impact on the scattering
amplitudes \cite{Lutz98,Oset00,Laura,Effenber00,Lutz02,lauran} and
thus on the in-medium properties of the antikaons.  Due to the
complexity of this many-body problem the actual kaon and antikaon
self energies (or potentials) are still a matter of debate.

The topic of isospin effects in asymmetric nuclear matter has
gained interest in the recent past \cite{asym}. The isospin
effects are important in isospin asymmetric heavy-ion collision
experiments.  Within the UrQMD model the density dependence of the
symmetry potential has been studied by investigating observables
like the $\pi^-/\pi^+$ ratio, the n/p ratio \cite{li1}, the
$\Delta ^-/\Delta ^{++}$ ratio as well as the effects on the
production of $K^0$ and $K^+$ \cite {li2} and on pion flow
\cite{li3} for neutron rich heavy ion collisions. Recently, the
isospin dependence of the in-medium NN cross section \cite{li4}
has also been studied.

In the present investigation we will use a chiral SU(3) model for
the description of hadrons in the medium \cite{paper3}.  The
nucleons -- as modified in the hot hyperonic matter -- have
previously been studied within this model \cite{kristof1}.
Furthermore, the properties of vector mesons
\cite{hartree,kristof1} -- modified by their interactions  with
nucleons in the medium -- have also been examined and have been
found to have sizeable modifications due to Dirac sea
polarization effects. The chiral SU(3)$_{flavor}$ model was
generalized to SU(4)$_{flavor}$ to study the mass modification of
D-mesons arising from their interactions with the light hadrons in
hot hadronic matter in \cite{dmeson}. The energies of kaons
(antikaons), as modified in the medium due to
their interaction with nucleons, consistent with the low energy KN
scattering data \cite{juergen,cohen}, were also studied within
this framework \cite{kmeson1,isoamss}. In the present work, we investigate the
effect of isospin asymmetry on the kaon and antikaon optical
potentials in the asymmetric nuclear matter, consistent with the
low energy kaon nucleon scattering lengths for channels I=0 and I=1.
The pion nucleon scattering lengths are also calculated.


The outline of the paper is as follows: In section II we shall
briefly review the SU(3) model used in the present investigation.
Section III describes the medium modification of the K($\bar K$)
mesons in this effective model. In section IV, we discuss the
results obtained for the optical potentials of the kaons and
antikaons and the isospin-dependent effects on these optical
potentials in asymmetric nuclear matter. Section V summarizes our results
and discusses possible
extensions of the calculations.

\section{ The hadronic chiral $SU(3) \times SU(3)$ model }
In this section the various terms of the effective hadronic Lagrangian
used
\be
{\cal L} = {\cal L}_{kin} + \sum_{ W =X,Y,V,{\cal A},u }{\cal L}_{BW}
          + {\cal L}_{vec} + {\cal L}_0 + {\cal L}_{SB}
\label{genlag} \ee are discussed. Eq. (\ref{genlag}) corresponds
to a relativistic quantum field theoretical model of baryons and
mesons adopting a nonlinear realization of chiral symmetry
\cite{weinberg,coleman,bardeen} and broken scale invariance (for
details see \cite{paper3,hartree,kristof1}) to describe strongly
interacting nuclear matter. The model was used successfully to
describe nuclear matter, finite nuclei, hypernuclei and neutron
stars. The Lagrangian contains the baryon octet, the spin-0 and
spin-1 meson multiplets as the elementary degrees of freedom. In
Eq. (\ref{genlag}), $ {\cal L}_{kin} $ is the kinetic energy term,
$  {\cal L}_{BW}  $ contains the baryon-meson interactions in
which the baryon-spin-0 meson interaction terms generate the
baryon masses. $ {\cal L}_{vec} $ describes the dynamical mass
generation of the vector mesons via couplings to the scalar fields
and contains additionally quartic self-interactions of the vector
fields.  ${\cal L}_0 $ contains the meson-meson interaction terms
inducing the spontaneous breaking of chiral symmetry as well as a
scale invariance breaking logarithmic potential. $ {\cal L}_{SB} $
describes the explicit chiral symmetry breaking.

The kinetic energy terms are given as
\bea
\label{kinetic}
{\cal L}_{kin} &=& i\Tr \overline{B} \gamma_{\mu} D^{\mu}B
                + \frac{1}{2} \Tr D_{\mu} X D^{\mu} X
+  \Tr (u_{\mu} X u^{\mu}X +X u_{\mu} u^{\mu} X)
                + \frac{1}{2}\Tr D_{\mu} Y D^{\mu} Y \nonumber \\
               &+&\frac {1}{2} D_{\mu} \chi D^{\mu} \chi
                - \frac{ 1 }{ 4 } \Tr
\left(\tilde V_{ \mu \nu } \tilde V^{\mu \nu }  \right)
- \frac{ 1 }{ 4 } \Tr \left(F_{ \mu \nu } F^{\mu \nu }  \right)
- \frac{ 1 }{ 4 } \Tr \left( {\cal A}_{ \mu \nu } {\cal A}^{\mu \nu }
 \right)\, .
\eea
In (\ref{kinetic}) $B$ is the baryon octet, $X$ the scalar meson
multiplet, $Y$ the pseudoscalar chiral singlet, $\tilde{V}^\mu$ (${\cal
A}^\mu$) the renormalised vector (axial vector) meson multiplet with
the field strength tensor
$\tilde{V}_{\mu\nu}=\pa_\mu\tilde{V}_\nu-\pa_\nu\tilde{V}_\mu$ $({\cal
A}_{\mu\nu}= \pa_\mu{\cal A}_\nu-\pa_\nu{\cal A}_\mu $), $F_{\mu\nu}$
is the field strength tensor of the photon and $\chi$
is the scalar, iso-scalar dilaton (glueball) -field.
In the above, $u_\mu= -\fr{i}{2}[u^\dg\pam u - u\pam u^\dg]$,
where $u=\exp\Bigg[\fr{i}{\sigma_0}\pi^a\lambda^a\gamma_5\Bigg]$
is the unitary transformation operator. The covariant derivative
of a field $\Phi\equiv B,X, Y, A_\mu, V_\mu$
reads $ D_\mu \Phi = \pam\Phi + [\Gamma_\mu,\Phi]$ and
$\Gamma_\mu=-\fr{i}{4}[u^\dg\pam u-\pam {u^\dg} u
+ u\pam u^\dg- \pam u u^\dg ]$.

The baryon -meson interaction for a general meson field $W$ has
the form \be {\cal L}_{BW} = -\sqrt{2}g_8^W
\left(\alpha_W[\ovl{B}{\cal O}BW]_F+ (1-\alpha_W) [\ovl{B} {\cal
O}B W]_D \right) - g_1^W \frac{1}{\sqrt{3}} \Tr(\ovl{B}{\cal O}
B)\Tr W  \, , \ee with $[\ovl{B}{\cal O}BW]_F:=\Tr(\ovl{B}{\cal
O}WB-\ovl{B}{\cal O}BW)$ and $[\ovl{B}{\cal O}BW]_D:=
\Tr(\ovl{B}{\cal O}WB+\ovl{B}{\cal O}BW) - \frac{2}{3}\Tr
(\ovl{B}{\cal O} B) \Tr W$. The different terms to be considered
are those for the interaction of baryons  with scalar mesons
($W=X, {\cal O}=1$), with vector mesons  ($W=\tilde V_{\mu}, {\cal
O}=\gamma_{\mu}$ for the vector and $W=\tilde V_{\mu \nu}, {\cal
O}=\sigma^{\mu \nu}$ for the tensor interaction), with axial
vector mesons ($W={\cal A}_\mu, {\cal O}=\gamma_\mu \gamma_5$) and
with pseudoscalar mesons ($W=u_{\mu},{\cal
O}=\gamma_{\mu}\gamma_5$), respectively. For the current
investigation the following interactions are relevant:
Baryon-scalar meson interactions generate the baryon masses
through coupling of the baryons to the non-strange $ \sigma (\sim
\langle\bar{u}u + \bar{d}d\rangle) $ and the strange $
\zeta(\sim\langle\bar{s}s\rangle) $ scalar quark condensates. The
parameters $ g_1^S$, $g_8^S $ and $\alpha_S$ are adjusted to fix
the baryon masses to their experimentally measured vacuum values.
It should be emphasized that the nucleon mass also depends on the
{\em strange condensate} $ \zeta $. For the special case of ideal
mixing ($\alpha_S=1$ and $g_1^S=\sqrt 6 g_8^S$) the nucleon mass
depends only on the non--strange quark condensate. In the present
investigation, the general case will be used to study hot and
strange hadronic matter \cite{kristof1}, which takes into account
the baryon coupling terms to both scalar fields ($\sigma$ and
$\zeta$) while summing over the baryonic tadpole diagrams to
investigate the effect from the baryonic Dirac sea in the
relativistic Hartree approximation \cite{kristof1}.

In analogy to the baryon-scalar meson coupling there exist two
independent baryon-vector meson interaction terms corresponding to
the $F$-type (antisymmetric) and $D$-type (symmetric) couplings.
Here we will use the antisymmetric coupling because, following the
universality principle  \cite{saku69} and the vector meson
dominance model, one can conclude that the symmetric coupling
should be small. We realize this by setting $\alpha_V=1$ for all
fits. Additionally we decouple the strange vector field $
\phi_\mu\sim\bar{s} \gamma_\mu s $ from the nucleon by setting $
g_1^V=\sqrt{6}g_8^V $. The remaining baryon-vector meson
interaction reads \be {\cal
L}_{BV}=-\sqrt{2}g_8^V\Big\{[\bar{B}\gamma_\mu
BV^\mu]_F+\Tr\big(\bar{B}\gamma_\mu B\big) \Tr V^\mu\Big\}\, . \ee

The Lagrangian describing the interaction for the scalar mesons, $X$,
and pseudoscalar singlet, $Y$, is given as \cite{paper3}
\bea
\label{cpot}
{\cal L}_0 &= &  -\frac{ 1 }{ 2 } k_0 \chi^2 I_2
     + k_1 (I_2)^2 + k_2 I_4 +2 k_3 \chi I_3,
\eea with $I_2= \Tr (X+iY)^2$, $I_3=\det (X+iY)$ and $I_4 = \Tr
(X+iY)^4$. In the above, $\chi$ is the scalar color singlet
glueball field. It is introduced in order to mimic the QCD trace
anomaly, i.e. the non-vanishing energy-momentum tensor
$\theta_\mu^\mu = (\beta_{QCD}/2g)\langle
G^a_{\mu\nu}G^{a,\mu\nu}\rangle$, where $G^a_{\mu\nu}$ is the
gluon field tensor. A scale breaking potential is introduced: \be
\label{lscale}
  {\cal L}_{\mathrm{scalebreak}}=- \frac{1}{4}\chi^4 \ln
   \frac{ \chi^4 }{ \chi_0^4}
 +\frac{\delta}{3}\chi^4 \ln \frac{I_3}{\det \langle X \rangle_0}
\ee
which allows for the identification of the $\chi$ field width the gluon
condensate $\theta_\mu^\mu=(1-\delta)\chi^4$.
Finally the term
${\cal L}_{\chi} = - k_4 \chi^4 $
generates a phenomenologically consistent finite vacuum expectation
value. The variation of
$\chi$ in the medium is rather small \cite{paper3}.
Hence we shall use the frozen glueball approximation i.e. set
$\chi$ to its vacuum value, $\chi_0$.

The Lagrangian for the vector meson interaction is written as
\bea
{\cal L}_{vec} &=&
    \fr{m_V^2}{2}\fr{\chi^2}{\chi_0^2}\Tr\big(\tilde{V}_\mu\tilde{V}^\mu\big)
+   \fr{\mu}{4}\Tr\big(\tilde{V}_{\mu\nu}\tilde{V}^{\mu\nu}X^2\big) 
+ \fr{\lambda_V}{12}\Big(\Tr\big(\tilde{V}^{\mu\nu}\big)\Big)^2 +
    2(\tilde{g}_4)^4\Tr\big(\tilde{V}_\mu\tilde{V}^\mu\big)^2  \, .
\eea
The vector meson fields, $\tilde{V}_\mu$ are related to the
renormalized fields by
$V_\mu = Z_V^{1/2}\tilde{V}_\mu$, with $V = \omega, \rho, \phi \, $.
The masses of $\omega,\rho$ and $\phi$ are fitted from $m_V, \mu$ and
$\lambda_V$.

The explicit symmetry breaking term is given as \cite{paper3}
\be
 {\cal L}_{SB}=\Tr A_p\left(u(X+iY)u+u^\dagger(X-iY)u^\dagger\right)
\label{esb-gl}
\ee
with $A_p=1/\sqrt{2}{\mathrm{diag}}(m_{\pi}^2 f_{\pi},m_\pi^2 f_\pi, 2 m_K^2 f_K
-m_{\pi}^2 f_\pi)$ and $m_{\pi}=139$ MeV, $m_K=498$ MeV. This
choice for $A_p$, together with the constraints
$\sigma_0=-f_\pi$, $\zeta_0=-\frac {1}{\sqrt 2} (2 f_K -f_\pi)$
on the VEV on the scalar condensates assure that
the PCAC-relations of the pion and kaon are fulfilled.
With $f_{\pi} = 93.3$~MeV and $f_K = 122$~MeV we obtain $|\sigma_0| =
93.3$~MeV and $|\zeta_0 |= 106.76$~MeV.

We proceed to study the hadronic properties in the chiral SU(3) model.
The Lagrangian density in the mean field approximation is given as
\begin{eqnarray}
{\cal L}_{BX}+{\cal L}_{BV} &=& -\sum_i\overline{\psi_{i}}\, [g_{i
\omega}\gamma_0 \omega + g_{i\phi}\gamma_0 \phi
+m_i^{\ast} ]\,\psi_{i} \\
{\cal L}_{vec} &=& \frac{1}{2}m_{\omega}^{2}\frac{\chi^2}{\chi_0^2}\omega^
2+g_4^4 \omega^4 +
\frac{1}{2}m_{\phi}^{2}\frac{\chi^2}{\chi_0^2}\phi^2+g_4^4
\left(\fr{Z_\phi}{Z_\omega}\right)^2\phi^4\\
{\cal L}_{dilaton} &=& - \frac{ 1 }{ 2 } k_0 \chi^2
(\sigma^2+\zeta^2+\delta^2) + k_1 (\sigma^2+\zeta^2+\delta^2)^2
     + k_2 ( \frac{ \sigma^4}{ 2 } + \frac{\delta^4}{2} + \zeta^4)
\nonumber \\
     &+& k_3 \chi (\sigma^2 - \delta^2) \zeta
- k_4 \chi^4 - \frac{1}{4}\chi^4 \ln \frac{ \chi^4 }{ \chi_0^4}
 +\frac{\delta}{3} \chi^4
\ln \frac{(\sigma^2-\delta^2)\zeta}{\sigma_0^2 \zeta_0}\\
{\cal L}_{SB} &=& -\left[m_{\pi}^2 f_{\pi}
\sigma
+ (\sqrt{2}m_K^2 f_K - \frac{ 1 }{ \sqrt{2} } m_{\pi}^2 f_{\pi})\zeta
\right],
\end{eqnarray}
where $m_i^* = -g_{\sigma i}{\sigma}-g_{\zeta i}{\zeta}-g_{\delta i}{\delta} $ is the
effective mass of the baryon of type $i$ ($i = N, \Si, \La, \Xi $).
In the above, ${\cal L}_{dilaton}$ is the combined contribution from the 
Lagrangian densities ${\cal L}_0$, ${\cal L}_{scalebreak}$ and ${\cal L}_\chi$
and, $g_4=\sqrt {Z_\omega} \tilde g_4$ is the
renormalized coupling for $\omega$-field. The thermodynamical
potential of the grand canonical ensemble $\Omega$ per unit volume
$V$ at given chemical potential $\mu$ and temperature $T$ can be
written as \bea \frac{\Omega}{V} &=& -{\cal L}_{vec} - {\cal L}_0
- {\cal L}_{SB} - {\cal V}_{vac} + \sum_i\frac{\gamma_i }{(2
\pi)^3} \int d^3k\, E^{\ast}_i(k)\Big(f_i(k)+\bar{f}_i(k)
\Big)\nonumber \\
&&- \sum_i\frac{\gamma_i }{(2 \pi)^3}\,\mu^{\ast}_i
\int d^3k\,\Big(f_i(k)-\bar{f}_i(k)\Big).
\label{OmegaV}
\eea
Here the the potential at $\rho=0$ has been subtracted
in order to get a vanishing vacuum energy. In (\ref{OmegaV}) $\gamma_i$
are the spin-isospin degeneracy factors.  The $f_i$ and $\bar{f}_i$ are
thermal distribution functions for the baryon of species $i$, given in
terms of the effective single particle energy, $E^\ast_i$, and chemical
potential, $\mu^\ast_i$, as
\bea
f_i(k) &=& \fr{1}{{\rm e}^{\beta (E^{\ast}_i(k)-\mu^{\ast}_i)}+1}\quad ,\quad
\bar{f}_i(k)=\fr{1}{{\rm e}^{\beta (E^{\ast}_i(k)+\mu^{\ast}_i)}+1}, \no
\eea
with $E^{\ast}_i(k) = \sqrt{k_i^2+{m^\ast_i}^2}$ and $ \mu^{\ast}_i
                = \mu_i-g_{i\omega}\omega$.
The mesonic field equations are determined by minimizing the
thermodynamical potential \cite{hartree,kristof1}.
They depend on
the scalar and vector densities
for the baryons at finite temperature
\bea
\rho^s_i = \gamma_i
\int \frac{d^3 k}{(2 \pi)^3} \,\frac{m_i^{\ast}}{E^{\ast}_i}\,
\left(f_i(k) + \bar{f}_i(k)\right) \, ; \;\;
\rho_i = \gamma_i \int \frac{d^3 k}{(2 \pi)^3}\,\left(f_i(k) -
\bar{f}_i(k)\right) \,.
\label{dens}
\eea
The energy density and the pressure are given as,
$\epsilon = \Omega/V+\mu_i\rho_i $+TS and $ p = -\Omega/V $.

\begin{figure}
\phantom{a}\hspace*{-2cm}
\psfig{file=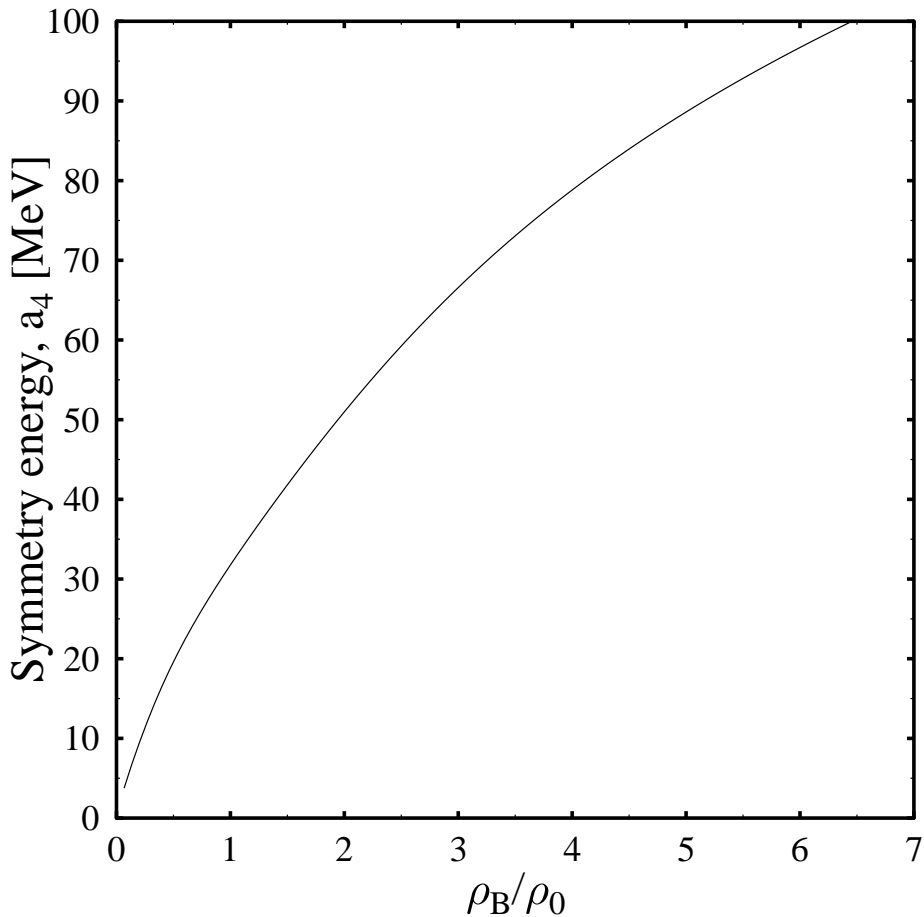,width=16cm}
\caption{
The symmetry energy, $a_4$ plotted
as a function of the baryon density, $\rho_B/\rho_0$.
}
\label{a4d2}
\end{figure}

\section{Kaon (antikaon) interactions in the chiral SU(3) model}
\label{kmeson}

In this section, we derive the dispersion relations
for the $K (\bar K)$ \cite{kmeson}  and calculate their optical potentials
in asymmetric nuclear matter \cite{isoamss}. The medium modified
energies of the kaons and antikaons arise from their interactions
with the nucleons and scalar mesons
within the chiral SU(3) model.

In this model,
the interactions of the kaons and antikaons to the scalar fields
(non-strange, $\sigma$ and strange,
$\zeta$), scalar--isovector field $\delta$ as well as a vectorial
interaction with the nucleons (the so--called Weinberg-Tomozawa interaction)
modify the energies for ${\rm K}(\bar {\rm  K})$ mesons in the medium.
In the following, we shall derive the dispersion relations
for the kaons and antikaons, including the effects from isospin
asymmetry originating from the scalar-isovector $\delta$ field
as well as vectorial interaction with the nucleons and a range term
arising from the interaction with the nucleons.
It might be noted here that the interaction of the pseudoscalar
mesons to the vector mesons, in addition to the pseudoscalar meson--nucleon
vectorial interaction leads to a double counting in the linear realization
of the chiral effective theory \cite{borasoy}. Within the
nonlinear realization of the chiral effective theories, such an interaction
does not arise in the leading or sub-leading order, but only as a higher
order contribution \cite{borasoy}. Hence the vector meson-pesudoscalar
interaction will not be considered within the present investigation.

\begin{figure}
\phantom{a}\hspace*{-2cm}
\psfig{file=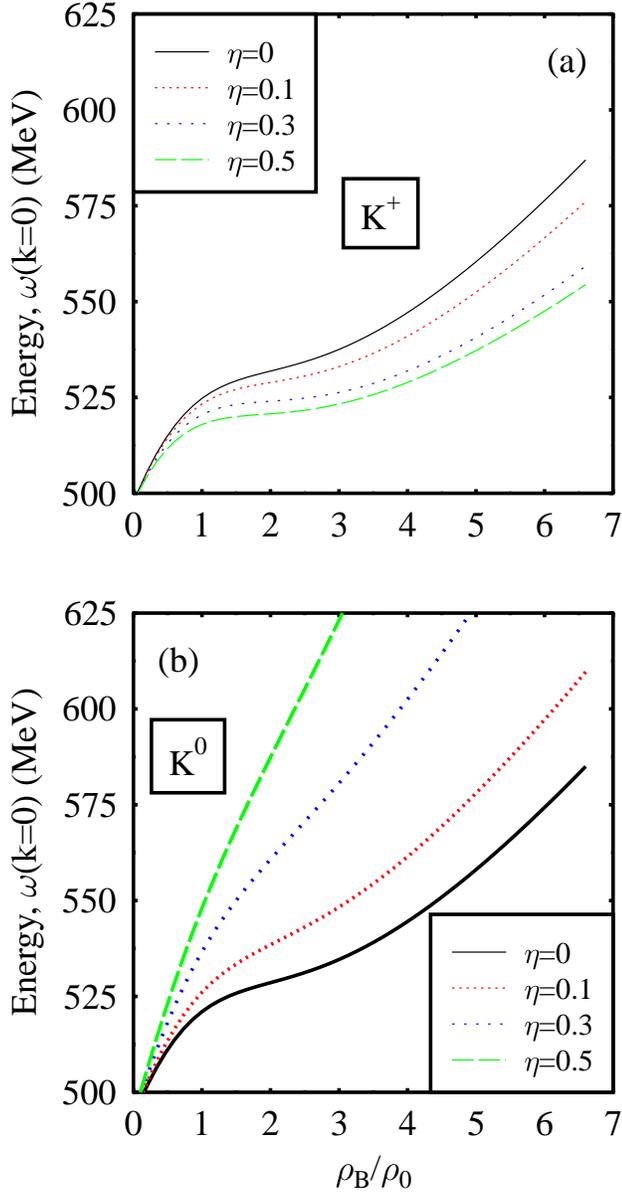,width=16cm}
\caption{
The kaon energies (for $K^+$ in (a) and for $K^0$ in (b)) in MeV plotted
as a functions of the baryon density, $\rho_B/\rho_0$
for different values of the isospin asymmetry  parameter, $\eta$.
}
\label{mkaon}
\end{figure}

\begin{figure}
\phantom{a}\hspace*{-2cm}
\psfig{file=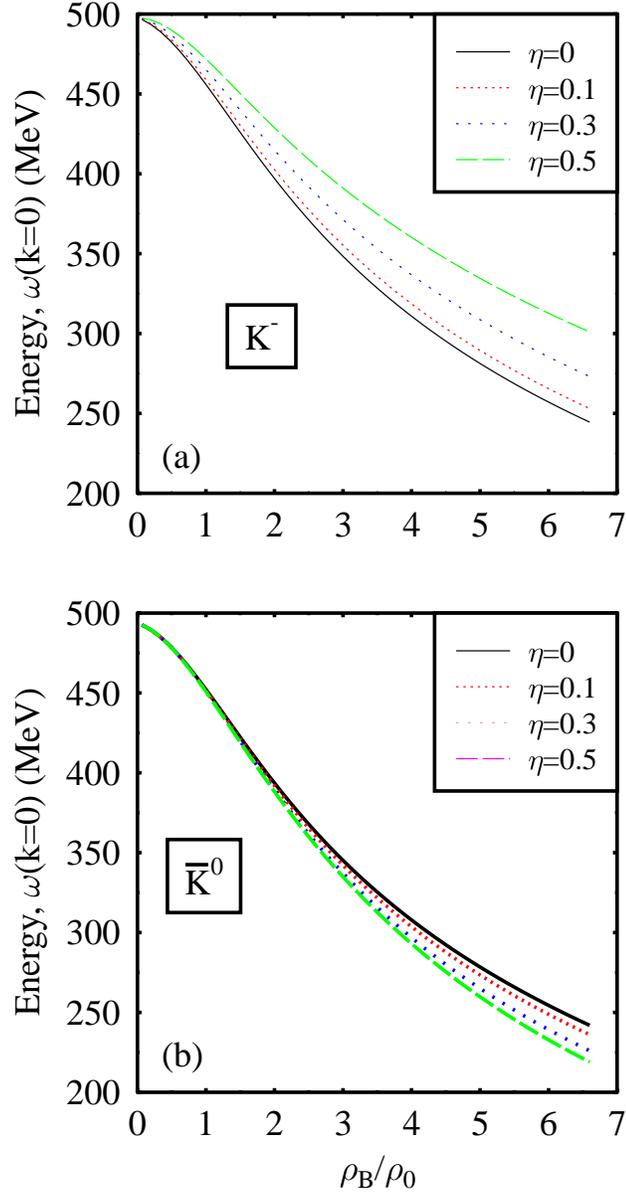,width=16cm}
\caption{
The energies of the antikaons ( for $K^-$ in (a) and for $\bar {K^0}$ in (b)),
at zero momentum as functions of the baryon density
($\rho_B/\rho_0$),
are plotted for different values of the isospin asymmetry parameter, $\eta$.
}
\label{mkbar}
\end{figure}

\begin{figure}
\phantom{a}\hspace*{-2cm}
\psfig{file=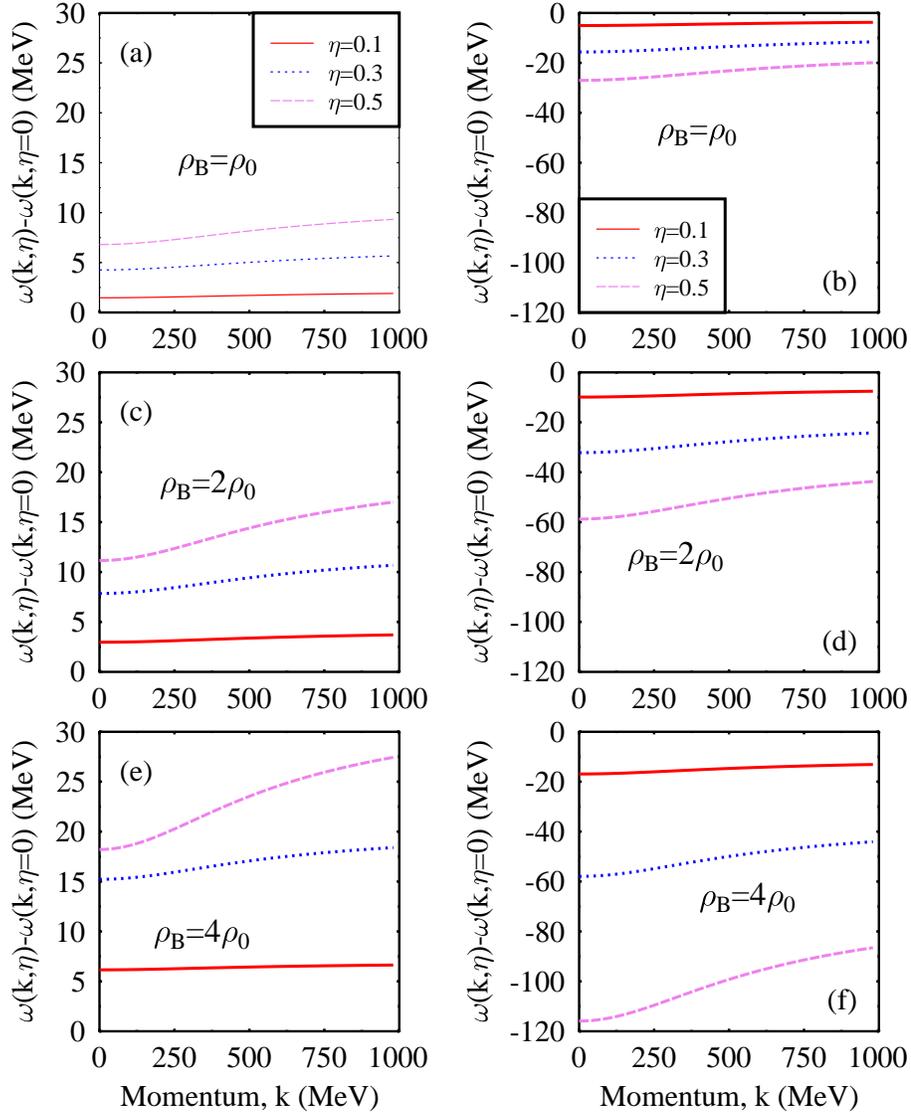,width=16cm}
\caption{
The kaon energies (for $K^+$ in (a), (c) amd (e) and for $K^0$
in (b), (d) and (f)) as compared to the isospin symmetric case, plotted
as a functions of the momentum for various values of the baryon density,
$\rho_B$ and
for different values of the isospin asymmetry  parameter, $\eta$.
}
\label{domgkaon}
\end{figure}

\begin{figure}
\phantom{a}\hspace*{-2cm}
\psfig{file=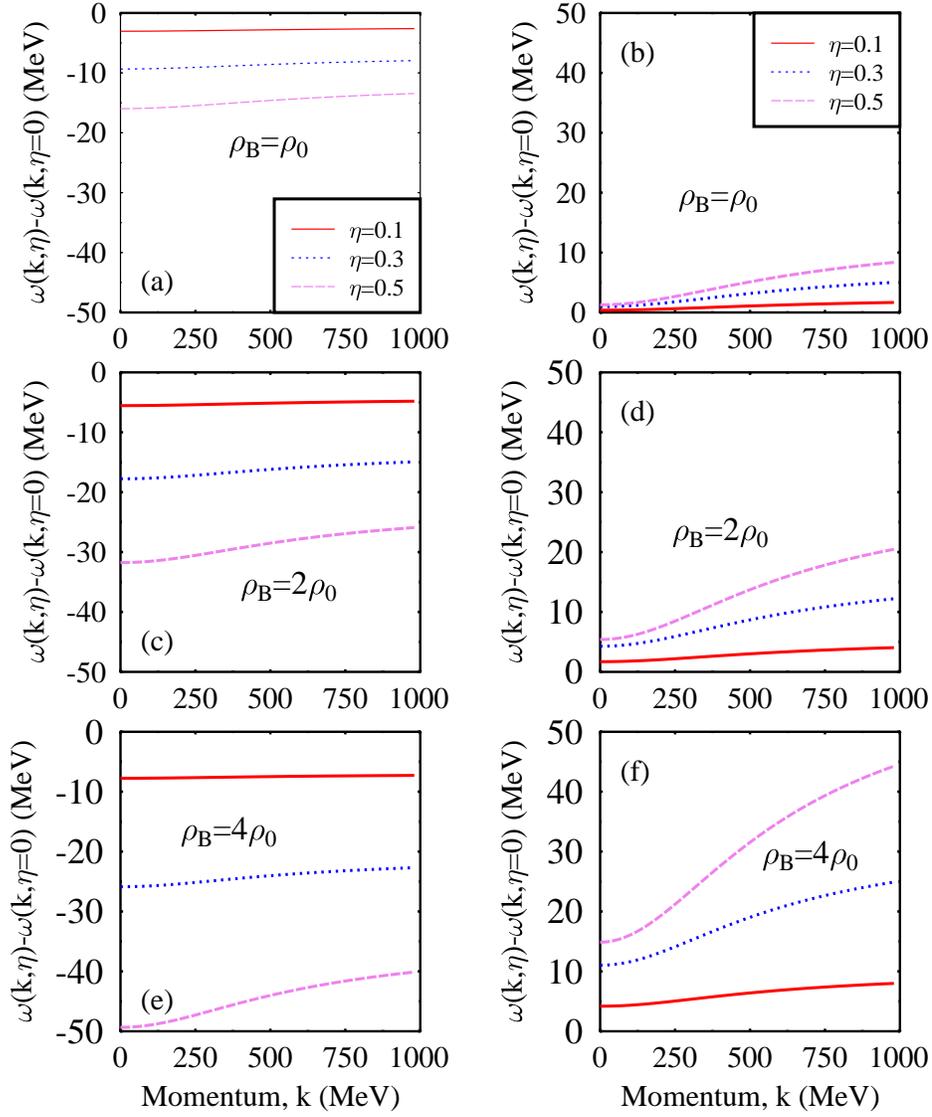,width=16cm}
\caption{
The antikaon energies (for $K^-$ in (a), (c) amd (e) and for $\bar {K^0}$
in (b), (d) and (f)) as compared to the isospin symmetric case, plotted
as a functions of the momentum,
for different values of the isospin asymmetry  parameter, $\eta$
for various values of the baryon density, $\rho_B$.
}
\label{omgkbar}
\end{figure}

\begin{figure}
\phantom{a}\hspace*{-2cm}
\psfig{file=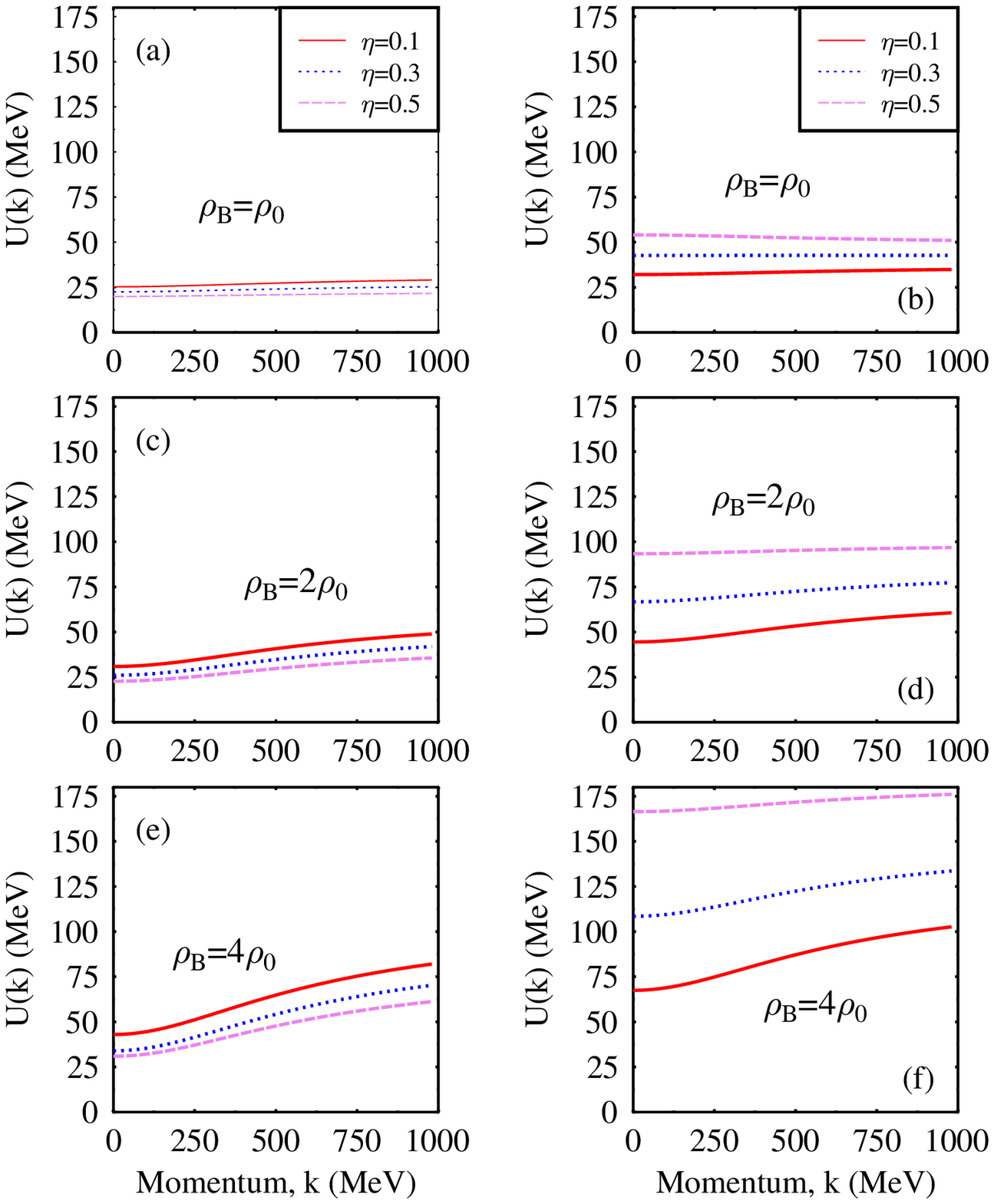,width=16cm}
\caption{
The kaon optical potentials (for $K^+$ in (a), (c) and (e) and for $K^0$
in (b), (d) and (f)) in MeV,
plotted as functions of the momentum for various baryon densities, $\rho_B$
and for different values of the isospin asymmetry  parameter, $\eta$.
}
\label{optkaonk}
\end{figure}

\begin{figure}
\phantom{a}\hspace*{-2cm}
\psfig{file=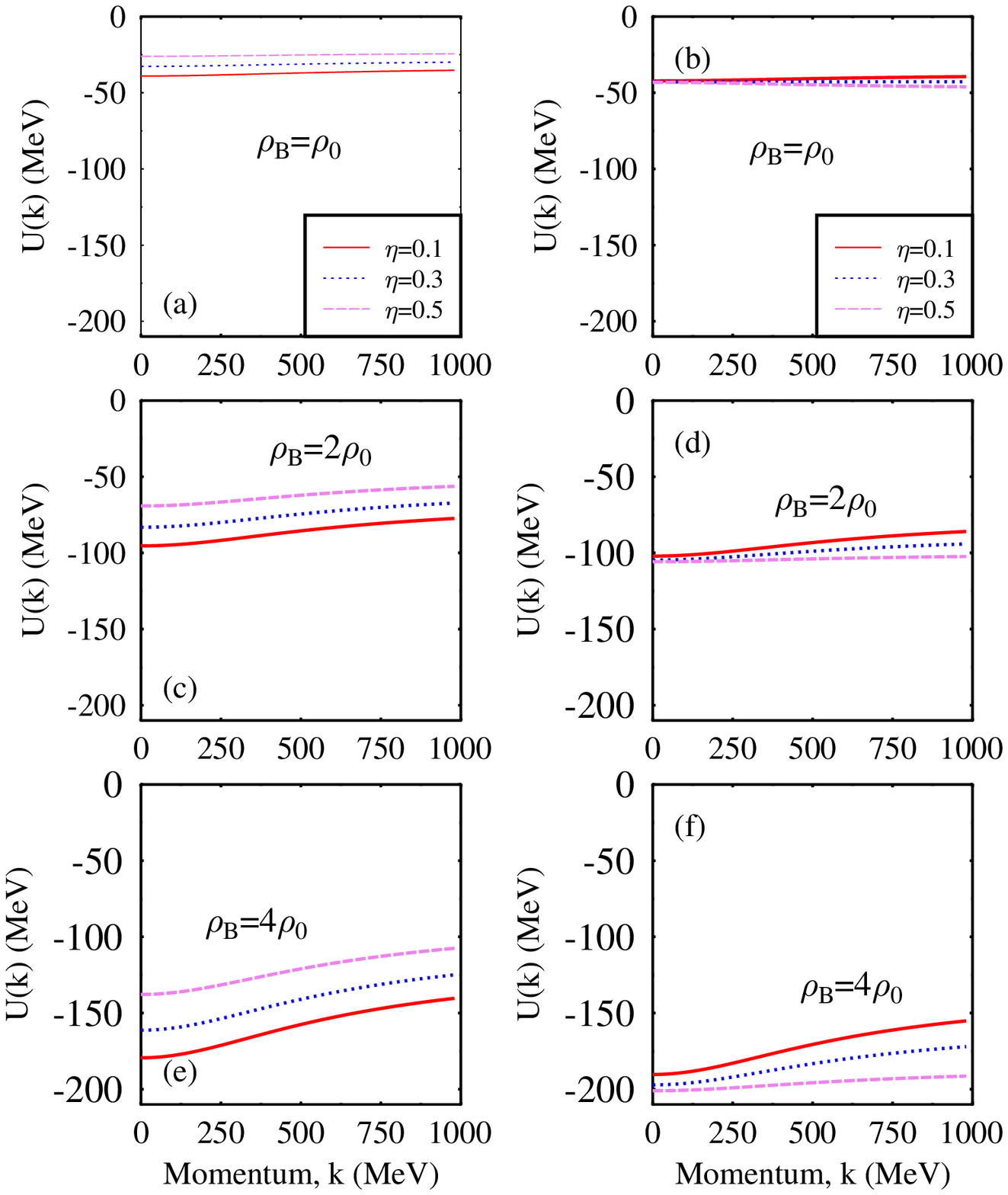,width=16cm}
\caption{
The antikaon optical potentials (for $K^-$ in (a), (c) and (e) and for
$\bar {K^0}$ in (b), (d) and (f)) in MeV,
plotted as functions of the momentum for various baryon densities, $\rho_B$
and for different values of the isospin asymmetry  parameter, $\eta$.
}
\label{optkbark}
\end{figure}

%
The scalar meson multiplet has the expectation value
$\langle X \rangle
= \rm {diag}( (\sigma+\delta)/\sqrt 2, (\sigma-\delta)/\sqrt 2 , \zeta ) $,
with $\sigma$ and $\zeta$ corresponding to the non-strange and strange
scalar condensates, and, $\delta$ is the third isospin component of the
scalar-isovector field, $\vec \delta$.
The pseudoscalar meson field $P$ can be written as,
\begin{equation}
P = \left(
\begin{array}{ccc} \pi^0/\sqrt 2 & \pi^+ & \frac{2 K^+}{1+w} \\
\pi^- & -\pi^0/\sqrt 2 & \frac {2 K^0}{1+w} \\
\frac {2 K^-}{1+w} & \frac {2 \bar {K^0}}{1+w} & 0
\\ \end{array}\right),
\end{equation}
where $w=\sqrt 2 \zeta/\sigma$ and we have written down
the terms that are relevant for the present investigation.
From PCAC one gets the decay constants for the pseudoscalar mesons
as $f_\pi=-\sigma$ and $f_K=-(\sigma +\sqrt 2 \zeta )/2$.

The scalar meson exchange interaction term
is determined from the explicit symmetry breaking term
by equation (\ref {esb-gl}), where $A_p =1/\sqrt 2$
diag ($m_\pi^2 f_\pi$, $m_\pi^2 f_\pi$, 2 $m_K^2 f_K -m_\pi^2 f_\pi$).

The interaction Lagrangian modifying the energies of the $K(\bar K)$-mesons
is given as
\begin{eqnarray}
\cal L _{KN} & = & -\frac {i}{8 f_K^2} \Big [ 3 ( \bar N \gamma^\mu N)
(\bar K (\partial_\mu K) - (\partial_\mu {\bar K})  K)
+(\bar N \gamma^\mu \tau^a N)
(\bar K \tau^a (\partial_\mu  K) - (\partial_\mu {\bar K})\tau ^a  K)\Big ]
\nonumber \\
 &+ & \frac{m_K^2}{2f_K} \Big [ (\sigma +\sqrt 2 \zeta)(\bar  K K)
+\delta ^a (\bar  K \tau^a K)\Big ] \nonumber \\
& - & \frac {1}{f_K}\Big [ (\sigma +\sqrt 2 \zeta)
(\partial _\mu {\bar  K})(\partial ^\mu K)
+(\partial _\mu  {\bar  K})\tau^a(\partial ^\mu K)\delta ^a \Big ]
\nonumber \\
&+ &\frac {d_1}{2 f_K^2}(\bar N N)
(\partial _\mu {\bar K})(\partial ^\mu K)\nonumber \\
&+& \frac {d_2}{2 f_K^2} \Big [
(\bar p p) (\partial_\mu K^+)(\partial^\mu K^-)
+(\bar n n) (\partial_\mu K^0)(\partial^\mu {\bar {K^0}})
\nonumber \\
&+&(\bar p n) (\partial_\mu K^+)(\partial^\mu {\bar {K^0}})
+(\bar n p) (\partial_\mu K^0)(\partial^\mu {\bar {K^-}})
\Big ]
\label{lagd}
\end{eqnarray}


In the above, $K$ and $\bar K$ are the kaon ($K^+$,$K^0$) and
antikaon ($K^-$,$\bar {K^0}$) doublets.
In (\ref{lagd}) the first line is the vectorial interaction term
obtained from the first term in (\ref{kinetic}) (Weinberg-Tomozawa
term).  The second term, which gives an attractive interaction for the
$K$-mesons, is obtained from the explicit symmetry breaking term
(\ref{esb-gl}).
The third term arises within the present chiral model from
the kinetic term of the pseudoscalar mesons given by the third term in
equation (\ref{kinetic}), when the scalar fields in one of the meson
multiplets, $X$, are replaced by their vacuum expectation values.  The
fourth and fifth terms in (\ref{lagd}) for the KN interactions arise
from the terms
\begin{equation}
{\cal L }_{(d_1)}^{BM} =d_1 Tr (u_\mu u ^\mu \bar B B),
\label{d1ld}
\end{equation}
and,
\begin{equation}
{\cal L }_{(d_2)}^{BM} =d_2 Tr (\bar B u_\mu u ^\mu B).
\label{d2ld}
\end{equation}

in the SU(3) chiral model \cite{kmeson1,isoamss}. The last three
terms in (\ref{lagd}) represent the range term in the chiral
model, with the last term being an isospin asymmetric interaction.
The Fourier transformation of the equation-of-motion for kaons
(antikaons) leads to the dispersion relations,
$$-\omega^2+ {\vec k}^2 + m_K^2 -\Pi(\omega, |\vec k|,\rho)=0,$$
where $\Pi$ denotes the kaon (antikaon) self energy in the medium.

Explicitly, the self energy $\Pi (\omega,|\vec k|)$ for the kaon doublet
arising from the interaction (\ref{lagd}) is given as

\begin{eqnarray}
\Pi (\omega, |\vec k|) &= & -\frac {1}{4 f_K^2}\Big [3 (\rho_p +\rho_n)
\pm (\rho_p -\rho_n)\Big ] \omega
+\frac {m_K^2}{2 f_K} (\sigma ' +\sqrt 2 \zeta ' \pm \delta ')
\nonumber \\ & +& \Big [- \frac {1}{f_K}
(\sigma ' +\sqrt 2 \zeta ' \pm \delta ')
+\frac {d_1}{2 f_K ^2} (\rho_s ^p +\rho_s ^n)\nonumber \\
&+&\frac {d_2}{4 f_K ^2} \Big ((\rho_s ^p +\rho_s ^n)
\pm   (\rho_s ^p -\rho_s ^n)\Big )
\Big ]
(\omega ^2 - {\vec k}^2),
\label{selfk}
\end{eqnarray}
where the $\pm$ signs refer to the $K^+$ and $K^0$ respectively.
In the above, $\sigma'(=\sigma-\sigma _0)$,
$\zeta'(=\zeta-\zeta_0)$ and  $\delta'(=\delta-\delta_0)$
are the fluctuations of the scalar-isoscalar fields $\sigma$ and $\zeta$,
and the third component of the scalar-isovector field, $\delta$,
from their vacuum expectation values.
The vacuum expectation value of $\delta$ is zero ($\delta_0$=0), since
a nonzero value for it will break the isospin symmetry of the vacuum
(the small isospin breaking effect coming from the mass and charge difference of the up and down quarks
has been neglected here).
$\rho_p$ and $\rho_n$ are the number densities
for the proton and the neutron, and $\rho_s^p$
and $\rho_s^n$ are their scalar densities.

Similarly, for the antikaon doublet, the self-energy is calculated as
\begin{eqnarray}
\Pi (\omega, |\vec k|) &= & \frac {1}{4 f_K^2}
\Big [ 3 (\rho_p +\rho_n)\pm (\rho_p -\rho_n)\Big ] \omega
+\frac {m_K^2}{2 f_K} (\sigma ' +\sqrt 2 \zeta ' \pm \delta ')
\nonumber \\ & +& \Big [- \frac {1}{f_K}
(\sigma ' +\sqrt 2 \zeta ' \pm \delta ')
+\frac {d_1}{2 f_K ^2} (\rho_s ^p +\rho_s ^n)\nonumber \\
&+&\frac {d_2}{4 f_K ^2}\Big ( (\rho_s ^p +\rho_s ^n)
\pm   (\rho_s ^p -\rho_s ^n)\Big )
 \Big ]
(\omega ^2 - {\vec k}^2),
\label{selfkb}
\end{eqnarray}
where the $\pm$ signs refer to the $K^-$ and $\bar {K^0}$ respectively.

After solving the above dispersion relations
for the kaons and antikaons, their optical potentials
can be calculated from
\be
U(\omega, k) = \omega (k) -\sqrt {k^2 + m_K ^2},
\ee
where $m_K$ is the vacuum mass for the kaon (antikaon).

The parameters $d_1$ and $d_2$ are calculated from the
empirical values of the KN scattering lengths
for I=0 and I=1 channels, given by
\bea
a_{KN}(I=0) & = & \frac {m_K}{4\pi {f_K}^2 \big (1+{m_K}/{m_N}\big )}
\Bigg [ -\frac {m_K f_K}{2}\bigg (\frac { g_{\sigma N}}{{m_\sigma}^2}
+\sqrt 2 \frac {  g_{\zeta N}}{{m_\zeta}^2}
- 3\frac {  g_{\delta N}}{{m_\delta}^2}\bigg )\nonumber \\
&+&\frac {(d_1 -d_2)m_K}{2}  \Bigg]
\label{akn0}
\eea

and

\bea
a_{KN}(I=1) & = & \frac {m_K}{4\pi {f_K}^2 \big (1+{m_K}/{m_N}\big )}
\Bigg [-1 -\frac {m_K f_K}{2}\bigg (\frac { g_{\sigma N}}{{m_\sigma}^2}
+\sqrt 2 \frac {  g_{\zeta N}}{{m_\zeta}^2}
+ \frac {  g_{\delta N}}{{m_\delta}^2}\bigg )\nonumber \\
&+&\frac {(d_1 +d_2)m_K}{2}  \Bigg]
\label{akn1}
\eea

These are taken to be
\cite{thorsson,juergen,barnes}
\be
a _{KN} (I=0) \approx -0.31 ~ {\rm {fm}},\;\;\;\;
a _{KN} (I=1) \approx -0.09 ~ {\rm {fm}}.
\label{akn01emp}
\ee
leading to the isospin averaged KN scattering length as
\be
\bar a _{KN}=\frac {1}{4} a_{KN}(I=0)+
\frac {3}{4} a_{KN}(I=1) \approx -0.255 ~ \rm {fm}.
\label{aknemp}
\ee

The pion nucleon scattering lengths given by
\be
a_{\pi N}\Bigg(I=\frac {3}{2}\Bigg)  =  \frac {m_\pi}{4\pi {f_\pi}^2
\big (1+({m_\pi}/{m_N})\big )}
\Bigg [-\frac {1}{2} -\frac {g_{\sigma N}}{{m_\sigma}^2} m_\pi f_\pi
+\frac {(d_1 +d_2)m_\pi}{2}  \Bigg]
\label{apn32}
\ee
and
\be
a_{\pi N}\Bigg(I=\frac {1}{2}\Bigg)  =  \frac {m_\pi}{4\pi {f_\pi}^2
\big (1+({m_\pi}/{m_N})\big )}
\Bigg [1 -\frac {g_{\sigma N}}{{m_\sigma}^2} m_\pi f_\pi
+\frac {(d_1 +d_2)m_\pi}{2}  \Bigg].
\label{apn12}
\ee
are also calculated in the present work.
\section{Results and Discussions}
\label{kmass}

The present calculations use the following model parameters. The
values, $g_{\sigma N}=10.6,\;\; {\rm and}\;\; g_{\zeta N}=-0.47$
are determined by fitting vacuum baryon masses. The other
parameters as fitted to the asymmetric nuclear matter saturation
properties in the mean field approximation are: $g_{\omega
N}$=13.3, $\g_{\rho N}$=5.5, $g_4$=79.7, $g_{N \delta}$=2.5,
$m_\zeta$ =1024.5 MeV, $m_\sigma$= 466.5 MeV and $m_\delta$=899.5
MeV. The coefficients $d_1$ and $d_2$, calculated from the
empirical values of the KN scattering lengths for I=0 and I=1
channels (\ref{akn01emp}), are $5.5/{m_K}$ and $0.66/{m_K}$
respectively. Using these parameters, the symmetry energy defined
as \be a_4 = 2 \frac{d^2 E}{d\eta ^2}|_{\eta=0} \ee with the
asymmetry parameter $\eta =\frac {1}{2} (\rho_n-\rho_p)/\rho_B$,
has a value of $a_4 =$ 31.7 MeV at saturation nuclear matter
density of $\rho_0$=0.15 fm$^{-3}$. Figure 1 shows the density
dependence of the symmetry energy, which increases with density
similar to previous calculations \cite{asymdens}.

The values of the pion nucleon scattering lengths
\cite{koch,arndt,pavan,beane,ericson,psi} are calculated
in the present model, with the values of $d_1$ and $d_2$ as
obtained by fitting the kaon nucleon scattering lengths. Their
values for $I=3/2$ and $I=1/2$, given by equations (\ref{apn32})
and (\ref{apn12}) are obtained as $a_{\pi N} (I=\frac{3}{2}) = -0.1474$ fm and $
a_{\pi N} (I=\frac{1}{2}) =
0.1823$ fm respectively. This determines the isoscalar and
isovector scattering lengths for $\pi N$ scattering
\big ($a_+=\big (a_{\pi N}(I=\frac{1}{2})+2a_{\pi N}(I=\frac{3}{2})\big )/3$
and $a_-=\big (a_{\pi N}(I=\frac{1}{2})-a_{\pi N}(I=\frac{3}{2})\big )/3$
\big)
to be $a_+=-0.0266/{m_\pi}$ and $a_-= 0.078/{m_\pi}$. These may be
compared with the results of $a_+=-0.0029/{m_\pi}$ and $a_-=0.0936/{m_\pi}$
derived from  pionic atoms \cite{beane}, the values $a_+=-0.0012/m_\pi$
and $a_-=0.0895/m_\pi$ using the empirical values of the $\pi^- p$ and
$\pi^- d$ scattering lengths \cite{ericson} and the values
$a_+=-0.0001/m_\pi$ and $a_-=0.0885/m_\pi$ from pionic deuterium shift
by the experimental PSI group \cite{psi}.

The $\pi N$ and $KN$ sigma terms are also calculated within the model.
They turn out to be 44 MeV and 725 MeV for the set of parameters
used in the present calculations.

The kaon and antikaon properties were studied
in the isospin symmetric hadronic matter
within the chiral SU(3) model in ref. \cite{kmeson1}.
The contribution from the vector interaction (Weinberg-Tomozawa
term) leads to a drop for the antikaon energy, whereas they
are repulsive for the kaons. The scalar meson
exchange term arising from the scalar-isoscalar fields ($\sigma$
and $\zeta$) is attractive for both $K$ and $\bar K$. The first
term of the range term of eq. (\ref{lagd}) is
repulsive whereas the second term has an attractive contribution
for the isospin symmetric matter \cite{kmeson1} for both kaons
and antikaons. The third term of the range term has an isospin
asymmetric contribution.

The contributions from (i) the last term of the Weinberg-Tomozawa
term, (ii) the scalar-isovector, $\delta$-field as well as (iii)
the $d_2$ term in the interaction Lagrangian given by equation
(\ref{lagd}) introduce an isotopic asymmetry in the K and $\bar
{\rm K}$-energies. For $\rho_n > \rho_p$, in the kaon sector,
$K^+$ ($K^0$) has negative (positive) contributions from $\delta$.
The $\delta$ contribution from the scalar exchange term is
positive (negative) for $K^+$ ($K^0$), whereas that arising from
the range term has the opposite sign and dominates over the former
contribution.

In figure 2, the energies of the $K^+$ and $K^0$ at zero momentum,
are plotted for different values of the isospin asymmetry
parameter, $\eta$, at various densities. For $\rho_B=\rho_0$ the
energy of $K^+$ is seen to drop by about 7 MeV at zero momentum
when $\eta$ changes from 0 to 0.5. On the other hand, the $K^0$
energy is seen to increase by about 27 MeV for $\eta$=0.5, from
the isospin symmetric case of $\eta$=0. The reason for this
opposite behavior for the  $K^+$ and $K^0$ on the isospin
asymmetry originates from the vectorial (Weinberg-Tomozawa),
$\delta$ meson contribution as well as from the isospin dependent
range term ($d_2$- term) contributions.
For $K^+$, the $\eta$-dependence of
the energy is seen to be less sensitive at higher densities,
whereas the energy of $K^0$  is seen to have a larger drop from
the $\eta$=0 case, as we increase the density. 

For the antikaons, the $K^-(\bar {K^0})$ energy at zero momentum
is seen to increase (drop) with $\eta$, as we increase the density
as seen in figure 3. The sensitivity of  the isospin asymmetry
dependence of the energies is seen to be larger for $K^-$ with
density, whereas it becomes smaller for $\bar {K^0}$ at high
densities. The mass drop as modified in the isospin asymmetry in
neutron star matter, will have relevance for the onset of antikaon
(${K^-}$ and ${\bar {K^0}}$) condensation.

The energies of the kaons and antikaons, with respect to the
isospin symmetric case, for different values of the isospin
asymmetric parameter, $\eta$ are plotted as functions of the
momentum in figures 4 and 5.  The energies of the kaons and antikaons
are plotted for densities $\rho_B=\rho_0$, $2\rho_0$ and $4 \rho_0$ 
in the same figures. These are seen to be more sensitive
to momentum as we increase the isospin parameter. The momentum
dependence turns out to be stronger for higher densities, and in
particular, the effect seems to be more significant for $K^0$ (as
compared to $K^+$) and $K^-$ (as compared to $\bar {K^0}$).

The qualitative behavior of the isospin asymmetry dependencies of
the energies of the kaons and antikaons are also reflected in
their optical potentials plotted in figure 6 for the kaons, and in
figure 7, for the antikaons, at selected densities. The different
behavior of the $K^+$ and $K^0$, as well as for the $K^-$ and
$\bar {K^0}$ optical potentials in the dense asymmetric nuclear
matter should be observed in their production as well as
propagation in isospin asymmetric heavy ion collisions. In particular
an experimental study of the $K^-/\bar{K}^0$ ratio (and its dependence
on the kaon momenta) might be a promising tool to investigate the
isospin effects discussed here.
The effects of the isospin asymmetric optical potentials could thus be
observed in nuclear collisions at the CBM experiment at the
proposed project FAIR at GSI, where experiments with neutron rich
beams are planned to be carried out.

\section{Summary}

To summarize, within a chiral SU(3) model we have investigated the
density dependence of the $K, \bar K$-meson optical potentials in
asymmetric nuclear matter, arising from the interactions with
nucleons and scalar mesons. The properties of the light
hadrons -- as studied in a SU(3) chiral model -- modify the $K
(\bar K)$-meson properties in the hadronic medium. The model with
parameters fitted to reproduce the properties of hadron masses in
vacuum, nuclear matter saturation properties and low energy KN
scattering data, takes into account all terms up to the next to
leading order arising in chiral perturbative expansion for the
interactions of $K (\bar K)$-mesons with baryons. The $\pi N$
scattering lengths are also calculated for the fitted set of model
parameters as $a_+=-0.0266/m_\pi$ and $a_-=0.078/m_\pi$ and have been
compared with the other results in the literature
\cite{beane,ericson,psi}. The $\pi N$ and $KN$ sigma coefficients
are calculated within the present chiral effective model and have values of
$\Sigma_{\pi N}$=44 MeV and $\Sigma _{KN}$= 725 MeV.

There is a significant density dependence of the isospin asymmetry
on the optical potentials of the kaons and antikaons. The results
can be used in heavy-ion simulations that include mean fields for
the propagation of mesons \cite{kmeson}. The different potentials
of kaons and antikaons can be particularly relevant for
neutron-rich heavy-ion beams at the CBM experiment at the future
project FAIR at GSI, Germany, as well as at the experiments at the
proposed Rare Isotope Accelerator (RIA) laboratory, USA.
The $K^-/\bar{K}^0$ ratio for different isospin of projectile
and target is a promising observable to study these effects.
Furthermore, the medium modification of antikaons due to isospin
aymmetry in dense matter can have important consequences,
for example on the onset of antikaon condensation in the bulk
charge neutral matter in neutron stars.
The effects of hyperons as well as finite temparatures on
optical potentials of kaons and antikaons
and their possible implications on the neutron star phenomenology
as well as heavy ion collision experiments are the intended topics
of future investigation.

\begin{acknowledgements}

We thank S. Kubis, C. Hanhart, S. Mallik, J. Reinhardt
for many fruitful discussions.
One of the authors (AM) is grateful to the Institut
f\"ur Theoretische Physik Frankfurt for the
warm hospitality where the present work was initiated.
AM acknowledges financial support from
Alexander von Humboldt stiftung. The use of the resources of the
Frankfurt Center for Scientific Computing (CSC) is additionally
gratefully acknowledged.
\end{acknowledgements}


\begin{thebibliography}{1}

\bibitem{kaplan}
    D. B. Kaplan and A. E. Nelson, Phys. Lett. B {\bf 175}, 57 (1986);
    A. E. Nelson and D. B. Kaplan, {\it ibid}, 192, 193 (1987).
\bibitem{FOPI}
    D. Best {\it et al.}, FOPI Collaboration,
    Nucl. Phys. A {\bf 625}, 307 (1997).
\bibitem{Laue99}
    F. Laue {\it et al.}, KaoS Collaboration,
    Phys. Rev. Lett. {\bf 82}, 1640 (1999).
\bibitem{kaosnew}
    M. Menzel {\it et al.}, KaoS Collaboration,
    Phys. Lett. B {\bf 495}, 26 (2000).
\bibitem{Sturm01}
    C. Sturm {\it et al.}, KaoS Collaboration,
    Phys. Rev. Lett. {\bf 86}, 39 (2001).
\bibitem{Forster02}
    A.  F\"orster {\it et al.}, KaoS Collaboration,
    J. Phys. G {\bf 28}, 2011 (2002).
\bibitem{lix}
    G. Q. Li, C.-H. Lee, and G. E. Brown,
    Nucl. Phys. A {\bf 625}, 372 (1997).
\bibitem{cmko}
    C. M. Ko, J. Phys. G {\bf 27}, 327 (2001).
\bibitem{Li2001}
    S. Pal, C. M. Ko, and Z.-W. Lin,
    Phys. Rev. C {\bf 64}, 042201 (2001).
\bibitem{Cass97}
    W. Cassing, E.L. Bratkovskaya, U. Mosel, S. Teis, and A. Sibirtsev,
    Nucl. Phys. A {\bf 614}, 415 (1997).
\bibitem{brat97}
    E. L. Bratkovskaya, W. Cassing, and U. Mosel,
    Nucl. Phys. A {\bf 622}, 593 (1997).
\bibitem{CB99}
    W. Cassing and E. L. Bratkovskaya, Phys. Rep. {\bf 308}, 65 (1999).
\bibitem{laura03}
    W. Cassing, L. Tol\'os, E. L. Bratkovskaya, and A.Ramos,
    Nucl. Phys. A {\bf 727}, 59 (2003).
\bibitem{Effenber00}
    J.  Schaffner-Bielich, V. Koch, and M. Effenberger,
    Nucl. Phys. A {\bf 669}, 153 (2000).
\bibitem{Aichelin}
    C. Hartnack, H. Oeschler, and J.  Aichelin,
    Phys. Rev. Lett. {\bf 90}, 102302 (2003).
\bibitem{Fuchs}
     C. Fuchs, Amand Faessler, E. Zabrodin, and Yu-Ming Zheng,
     Phys. Rev. Lett. {\bf 86}, 1974 (2001);
     C.~Fuchs, nucl-th/0312052.
\bibitem {brown}
    G. E. Brown and M. Rho, Phys. Rev. Lett. {\bf 66}, 2720 (1991).
\bibitem{FGB94}
    E. Friedman, A. Gal, and C.J. Batty,
    Nucl. Phys. A {\bf 579}, 518  (1994).
\bibitem{Gal}
    A. Gal, Nucl. Phys. A {\bf 691}, 268 (2001).
\bibitem{Lutz98}
    M. Lutz, Phys. Lett. B {\bf 426}, 12 (1998).
\bibitem{Lutz021}
    M. Lutz and E. E. Kolomeitsev,
    Nucl.  Phys.  A {\bf 700}, 193 (2002).
\bibitem{Lutz02}
    M. Lutz and C. L. Korpa, Nucl. Phys. A {\bf 700}, 309 (2002).
\bibitem{Oset00}
    A. Ramos and E. Oset, Nucl. Phys. A {\bf 671}, 481 (2000).
\bibitem{lauran}
    L. Tol\'os, A.  Ramos, and A. Polls,
    Phys. Rev. C {\bf 65}, 054907 (2002).
\bibitem{Laura}
    L. Tol\'os, A. Ramos, A. Polls, and T.T.S. Kuo,
    Nucl.  Phys. A {\bf 690}, 547 (2001).
\bibitem {asym} B.A. Li, Phys. Rev. Lett. 88, 192701 (2002);
B.A.Li, G.C. Yong and W. Zuo, Phys. Rev. C {\bf 71}, 014608 (2005);
L. W. Chen, C. M. Ko and B. A. Li, Phys. Rev. C {\bf 72}, 064606 (2005).
T. Gaitanos, M. Di Toro, S. Typee, V. Baran, C. Fuchs, V. Greco,
H.H.Wolter, Nucl. Phys. A {\bf 732}, 24 (2004);
T. Gaitanos, M. Di Toro, G. Ferini, M. Colonna, H. H. Wolter,
nucl-th/0402041; Q. Li, Z. Li, E. Zhao, R. K. Gupta,
Phys. Rev. C {\bf 71}, 054907 (2005); A. Akmal and V. R. Pandharipandhe,
Phys. Rev. C {\bf 56}, 2261 (1997); H. Heiselberg and H. Hjorth-Jensen,
Phys. Rep. {\bf 328}, 237 (2000).
\bibitem {li1} Q. Li, Z. Li, S. Soff, M. Bleicher, H. St\"ocker,
Phys. Rev. C {\bf 72}, 034613 (2005)
\bibitem {li2} Q. Li, Z. Li, S. Soff, R. K. Gupta, M. Bleicher,
H. St\"ocker, J. Phys. G {\bf 31}, 1359 (2005).
\bibitem {li3} Q. Li, Z. Li, S. Soff, M. Bleicher,
H. St\"ocker, J. Phys. G {\bf 32}, 151 (2006).
\bibitem {li4} Q. Li, Z. Li, S. Soff, M. Bleicher,
H. St\"ocker, J. Phys. G {\bf 32}, 407 (2006).
\bibitem{paper3}
    P. Papazoglou, D. Zschiesche, S. Schramm, J. Schaffner-Bielich,
    H. St\"ocker, and W. Greiner, Phys. Rev. C {\bf 59},  411  (1999).
\bibitem{kristof1}
    A. Mishra, K. Balazs, D. Zschiesche, S. Schramm, H. St\"ocker,
    and W. Greiner, Phys. Rev. C {\bf 69}, 024903 (2004).
\bibitem{hartree}
    D. Zschiesche, A. Mishra, S. Schramm, H. St\"ocker, and W. Greiner,
    Phys. Rev. C {\bf 70}, 045202 (2004).
\bibitem{dmeson}
    A. Mishra, E. Bratkovskaya, J. Schaffner-Bielich,
    S. Schramm, and H. St\"ocker,
    Phys. Rev. C {\bf 69}, 015202 (2004).
\bibitem{juergen}
    J. Schaffner-Bielich, I. N. Mishustin, J. Bondorf,
    Nucl. Phys. A {\bf 625}, 325 (1997).
\bibitem{cohen}
J. Cohen, Phys. Rev. C {\bf 39}, 2285 (1989).
\bibitem{kmeson1}
     A. Mishra, E. L. Bratkovskaya, J. Schaffner-Bielich, S. Schramm
     and H. St\"ocker, Phys. Rev. C {\bf 70}, 044904 (2004).
\bibitem{isoamss}
A. Mishra and S. Schramm, Phys. Rev. C {\bf 74}, 064904 (2006).
\bibitem{weinberg}
S.Weinberg, Phys. Rev. {\bf 166} 1568 (1968).
\bibitem{coleman}
S. Coleman, J. Wess, B. Zumino, Phys. Rev. {\bf 177} 2239 (1969);
C.G. Callan, S. Coleman, J. Wess, B. Zumino, Phys. Rev. {\bf 177}
2247 (1969).
\bibitem{bardeen}
W. A. Bardeen and B. W. Lee, Phys. Rev. {\bf 177} 2389 (1969).
\bibitem{saku69}
    J.~J. Sakurai, {\em  Currents and Mesons},
    University of Chicago Press, Chicago, 1969.
\bibitem{kmeson}
    G. Mao, P. Papazoglou, S. Hofmann, S. Schramm,
    H. St\"ocker, and W. Greiner, Phys. Rev. C {\bf 59}, 3381 (1999).
\bibitem {borasoy}
B.Borasoy and U-G. Meissner, Int. J. Mod. Phys. A {\bf 11} 5183 (1996).
\bibitem{thorsson}
    G. E. Brown, C.-H. Lee, M. Rho, and V. Thorsson,
    Nucl. Phys. A {\bf 567}, 937 (1994).
\bibitem{barnes}
    T. Barnes and E. S. Swanson, Phys. Rev. C {\bf 49}, 1166 (1994).
\bibitem{asymdens}
    G-C Yong, B. A. Li and L. W. Chen, Phys. Rev. C {\bf 74}, 064617
(2006).
\bibitem{koch} R. Koch and E. Pietarinen, Nucl. Phys. {\bf A 336},
331 (1980).
\bibitem{arndt}
R. A. Arndt, M.M. Pavan, R. L. Workman and I. I. Strakovsky,
Scattering Interactive Dial-Up (SAID), VPI, Blacksburg, The VPI/GWU
$\pi$N solution SM99, 1999;
\bibitem{pavan}
M.M. Pavan, R. A. Arndt,  I.I. Strakovsky and R. L. Workman, $\pi$N
Newslett. {\bf 15}, 118 (1999).
\bibitem{beane}
      S.R. Beane, V. Bernard, E. Epelbaum, U-G. Meissner, D. R. Phillips,
Nuclear Physics A {\bf 720} 399 (2003).
\bibitem{ericson} T. E. O. Ericson, B. Loiseau and A. W. Thomas,
Phys. Rev. {\bf C 66}, 014005 (2002).
\bibitem{psi} H. Ch. Schroeder et al, Eur. Phys. J. C {\bf 21}, 473 (2001).

\end{thebibliography}
\end{document}